\documentclass[preprint]{aastex}

\newcommand{\masr}{mas yr$^{-1}$}

\newcommand{\jb}{Jy beam$^{-1}$}
\newcommand{\kms}{km s$^{-1}$}

\slugcomment
{Accepted on 21 March 2001 for publication in \\ the Astrophysical Journal\\~}

\shortauthors{Walker et al.}
\shorttitle{Motions in 3C~120}

\begin{document}


\title{THE STRUCTURE AND MOTIONS OF THE 3C~120 \\ RADIO JET ON SCALES OF
0.6 TO 300 PARSECS}

\author{R. C. Walker, J. M. Benson}
\affil{National Radio Astronomy Observatory\altaffilmark{1}, Socorro, NM 87801}
\email{cwalker@nrao.edu, jbenson@nrao.edu}

\author{S. C. Unwin}
\affil{Jet Propulsion Laboratory, Pasadena, CA, 91109}
\email{unwin@huey.jpl.nasa.gov}

\author{M. B. Lystrup\altaffilmark{2}}
\affil{Portland State University, Portland, OR, 97207}
\email{lystrup@physics.ucdavis.edu}

\author{T. R. Hunter}
\affil{Harvard-Smithsonian Center for Astrophysics, Cambridge, MA, 02138}
\email{thunter@cfa.harvard.edu}

\author{G. Pilbratt}
\affil{ESA Astrophysics Division/Space Science Department, Noordwijk, 
The Netherlands}
\email{gpilbratt@astro.estec.esa.nl}

\author{P. E. Hardee}
\affil{Department of Physics \& Astronomy, The University of Alabama,
Tuscaloosa, AL 35487}
\email{hardee@athena.astr.ua.edu}


\author{}
\affil{}
\email{}

\altaffiltext{1}{The National
Radio Astronomy Observatory is a facility of the National Science
Foundation, operated under cooperative agreement by Associated
Universities, Inc.}

\altaffiltext{2}{Current address: University of California Davis, Davis, 
California, 95616}

\begin{abstract}

Results are presented from long term VLBI monitoring of the parsec-scale 
radio jet in 3C~120 at 5 and 1.7 GHz.  The 5 GHz sequence includes
a few early epochs at 10.7 GHz.
Superluminal features are seen leaving the core at intervals of less
than a year, about as often as new features could be distinguished
with the 0.6 parsec resolution of the observations.  The underlying
jet is continuous, but not smooth, and the measured features are
simply the bright points in the convolution of the observing beam with
brightness fluctuations that occur on many scales.  The velocity of
different features varies, but not by more than about a factor of 2.
Clear variations in the velocity of an individual feature are not
seen.  Some features that were observed leaving the core in the 5 GHz
observations with 0.6 pc resolution are followed at 1.7 GHz, with 2.4
pc resolution, to projected distances in excess of 25 pc from the
core.  Older features up to at least 150 pc in projection from the
core are still moving at superluminal apparent speeds and are
therefore presumed to be relativistic.  Beyond that, the data are
inadequate for motion measurements.  The region where the jet slows to
non-relativistic speeds has not been found.

There are suggestions of stationary features, or brightening and
dimming regions, through which the moving features pass.  These may be
locations where there is interaction with the external medium, or they
may simply be the result of variations in the jet angle to the
line-of-sight.  The observation of stationary features in an otherwise
moving jet reinforces the idea that the lack of motion of the knot at
4\arcsec\ (2 kpc), that has been found not to be superluminal in other
observations, might not actually imply that the jet has slowed by that
position.  The structure of the jet in the vicinity of the most likely
stationary feature is suggestive of a helical pattern seen in
projection.  The deprojected wavelength of the pattern is very long
relative to the jet radius, unlike the situation in sources such as
M87 \citep{O89}.  If the 3C~120 jet does contain a slowly-moving,
helical structure, then theory suggests that the jet resides in a
relatively cool medium, not in a relativistically hot cocoon or lobe.

\end{abstract}


\section{Introduction}

The radio source 3C~120 exhibits structure on all scales from
under a parsec to hundreds of kiloparsecs \citep{W87}.  It is
dominated by a variable core that is resolved with Very Long Baseline
Interferometry (VLBI).  A prominent one-sided jet is seen extending
from the core on subparsec scales to about a hundred kiloparsecs.
On the largest scales there is a complex, two-sided lobe structure
extending to about 14\arcmin, or about half a megaparsec for
$z=0.033$ \citep{B80} and $H_0=75$ \kms Mpc$^{-1}$, the value that
will be used in this paper (this gives a scale of 0.60 pc mas$^{-1}$).
Superluminal motion is observed in the jet which, because of the low
redshift, can be studied with greater linear resolution than most
other extragalactic superluminal sources.  This was one of the first 4
sources in which superluminal motions were found in the early 1970s
and it has been studied with VLBI extensively ever since \citep[and
references therein]{S79,W82,B88,G00}.  The standard model for
superluminal motion involves relativistic motions in a jet oriented
nearly along the line-of-sight \citep{B79}.

The central region of the 3C~120 host galaxy is a strong and variable
emitter of radiation at all observed frequencies.  Because of the
presence of broad emission lines, the galaxy is usually classified as
a Seyfert~1 \citep{B67}, although its properties are also consistent
with those of a broad line radio galaxy and, to some extent, a low
luminosity quasar.  The galaxy is of order an arcminute in size, with
extensive H~II regions that may be ionized by the nucleus \citep{B80}.
Its morphology does not clearly match either a spiral or an
elliptical, and may be the disturbed product of a merger \citep{M88}.
\citet{MA91} give broadband spectra from infrared to X-rays and report
on monitoring observations over much of this range.  There is a
pronounced and variable UV excess which they model with a face-on
accretion disk that feeds a $4 \times 10^7 M_{\sun}$ black hole at a
variable rate of somewhat under a solar mass per year.  At optical
wavelengths, variability is observed in both the continuum
\citep{L79,W79,P79} and the spectral lines \citep{O80,P98}.
Reverberation mapping techniques have been used to derive a mass for
the central black hole of about $3 \times 10^7 M_{\sun}$ \citep{WP99}.
3C~120 is a complex and variable X-ray source with the unusual
property that the spectral index varies with the intensity
\citep{H85,MA91}.  \citet{GR97} find X-ray properties similar to those
of a radio quiet Seyfert, including the presence of strong iron lines,
indicating that beamed emission does not dominate at these
frequencies.

The radio jet is among the few that is seen at other wavelengths.  An
optical counterpart, extending out to 15\arcsec, was reported by
\citet{H95}.  The radio-to-optical spectral index is 0.65, about the
same as the radio spectral index on the same scales.  In addition,
there is an X-ray emission knot that is coincident with the 20\arcsec\
radio knot and is probably explained by synchrotron emission
\citep{H99}.


Recent high frequency VLBI observations at 86, 43 and 22 GHz
\citep{G98,G99,G00} have produced high resolution images and a
16-month movie of the innermost features of the jet.  The 86 GHz
Coordinated Millimeter VLBI Array observations put an upper limit on
the core size of $0.033$ pc; the core size lies below the resolution
of the 86 GHz image which is $54 \mu$as.  The 43 and 22 GHz
observations, made with the National Radio Astronomy Observatory's
(NRAO) Very Long Baseline Array (VLBA), show that the inner jet
structure is complex and exhibits superluminal motions on very small
scales.  The movie shows evidence for interaction with a cloud of
density and core distance intermediate between typical broad and
narrow line clouds.  The complexity of the structure at high
resolution makes it clear that features in lower resolution
observations, such as those presented in this paper, are likely to be
blends of smaller scale components.  The 43 and 22 GHz observations
show that the linear polarization on small scales is variable with
some features near the core having magnetic fields across the jet.
Polarization percentages of up to 10\% are seen, increasing with core
distance.  Farther from the core, and at much larger scales observed
with the VLA \citep{W87}, the polarization indicates that the magnetic
fields are parallel to the jet.  VLBA measurements made in 1996 showed
upper limits to circular polarization of about 0.2\% \citep{HW99}.

Recent studies have begun to apply physical models of jets to
observations of 3C~120.  The jet shows a variety of knots and
side-to-side structures, some reminiscent of helical structures.
Hydrodynamical jet theories incorporating shocks have been used to
model the moving knots in superluminal sources \citep[and references
therein]{G97}.  It has been suggested that the motion of the inner jet
structure 3C~120 may be described in this way \citep{G98}.  Both
moving and stationary features can be accommodated in such models.
Helical patterns are expected for some types of jet instabilities
\citep[and references therein]{H87,H00}.  The superluminal motion along
a curved trajectory seen in the radio source 3C~345 can be explained
by helical jet models \citep{H87,O89,S95}.

According to the standard model, the observation of superluminal
motions is good evidence that a jet is relativistic.  The fact that a
large proportion of bright, compact radio sources show superluminal
motions suggests that many jets are relativistic on parsec scales,
although there are selection effects due to relativistic beaming.  It
is not clear how far from the core the jets remain relativistic.  At
larger core distances, they become broader and features are larger, so
measurement of small angular velocities is more difficult.  Because of
its close proximity, 3C~120 has high angular rates of motion compared
to other superluminal sources making it a prime candidate for attempts
to find motions at large angular distances from the core.  So far
superluminal motion has been seen in the source to distances of tens
of parsecs \citep{B88}.  On arc second scales, a radio knot in the jet
4\arcsec\ from the core has been shown to have an upper limit on its
motions of significantly smaller than the speed of light
\citep{MU91,W97}.  This might suggest that the jet has slowed, or
moved out of the line-of-sight, by that distance.  But the knot could
be a stationary feature, perhaps a shock or bend, in a jet that is
still relativistic.  This possibility is supported by the fact that
the large scale structure is two sided, but the jet is still very one
sided on the scale of the knot, suggesting that the emission is still
beamed on that scale.  It is also supported by optical line
observations of regions on both sides of the nucleus that show
evidence of interaction with the jet \citep{H88,A89}.  One of the
goals of the 1.7 GHz observations presented here is to look for
motions on scales intermediate between those usually probed with VLBI
and the larger scale structures studied with linked interferometers.

In this paper we present the results obtained from five epochs of 1.7
GHz ($\lambda$=18 cm) VLBI observations spanning from 1982 to 1997.
We also present 5 GHz ($\lambda$=6 cm) VLBI monitoring data which was
taken approximately three times per year from 1977 through 1988.  A few
of the early observations of this sequence were made at 10.7 GHz.  With
this long series of observations, we can attempt to discern how
frequently new components appear, what sort of trajectory they follow,
and whether the velocities are constant for a given feature and
between features.  At 1.7 GHz, the core is not as dominant as at
higher frequencies and a wealth of jet structure is observable up
to about half an arcsecond from the core.

\section{The Observations}

\subsection{Monitoring at 5 GHz}

    Between 1977 and 1988, 3C~120 was observed approximately 3 times
per year with VLBI.  Until 1981, the observations alternated between
10.7 GHz and 5 GHz.  After 1981, it was realized that the motions in
the source were sufficiently fast that the interval between
same-frequency observations was too long to follow features
unambiguously.  After that date, nearly all observations were made at
5 GHz.  For this paper, the 10.7 GHz images have been convolved to
the same resolution as the 5 GHz observations and used as part of
the 5 GHz monitoring sequence.

    The early observations used the main antennas of the U.S. VLBI
Network.  As the years progressed, so did the size of the arrays as
more U.S. antennas and, especially, as European antennas were added.
The first observations involved only 3 antennas.  By the time the
series was terminated, a typical observation involved 11 antennas.
All of these observations used the Mark II VLBI recording system, with
a 2 MHz bandwidth and 1 bit sampling \citep{C73}.  Correlation was
done on one of the NRAO 3 station correlator, the Caltech 5 station
correlator, or the Caltech Block II (16 Mark II stations) correlator,
depending on epoch.  Results up to 1980.27 were reported by
\citet{W82}.  This paper will not consider the first few observations
because of their low quality, but will include observations after
1978.91 in the analysis, even though the first few of those were
previously reported.  The antennas involved in both these observations
and in the 1.7 GHz observations described below are listed in
Table~\ref{antennas}.  The date, frequency, antennas, and image peak
flux density per beam for each of the observations at 5 or 10.7 GHz
are listed in Table~\ref{obs6}.

\placetable{antennas}

\placetable{obs6}

    The 5 and 10.7 GHz observations were calibrated and imaged using
standard procedures using either the Caltech VLBI package or AIPS at
near the time of the observations.  The flux scales are usually based
on antenna temperatures measured at one or more of Effelsberg, Green
Bank, and Owens Valley.  They are probably good to about 10\% in
most cases.  However at least one image (1983.25) seems to have an
inconsistent flux density scale relative to the others; the
measured fluxes are low.  Perhaps this is because that epoch did not
include any short baselines.  No attempt at correction has been made,
partly for lack of an obvious way to determine a correction.  But
the dip in fluxes at that time should not be believed.

\begin{figure}
\epsscale{0.55}
\plotone{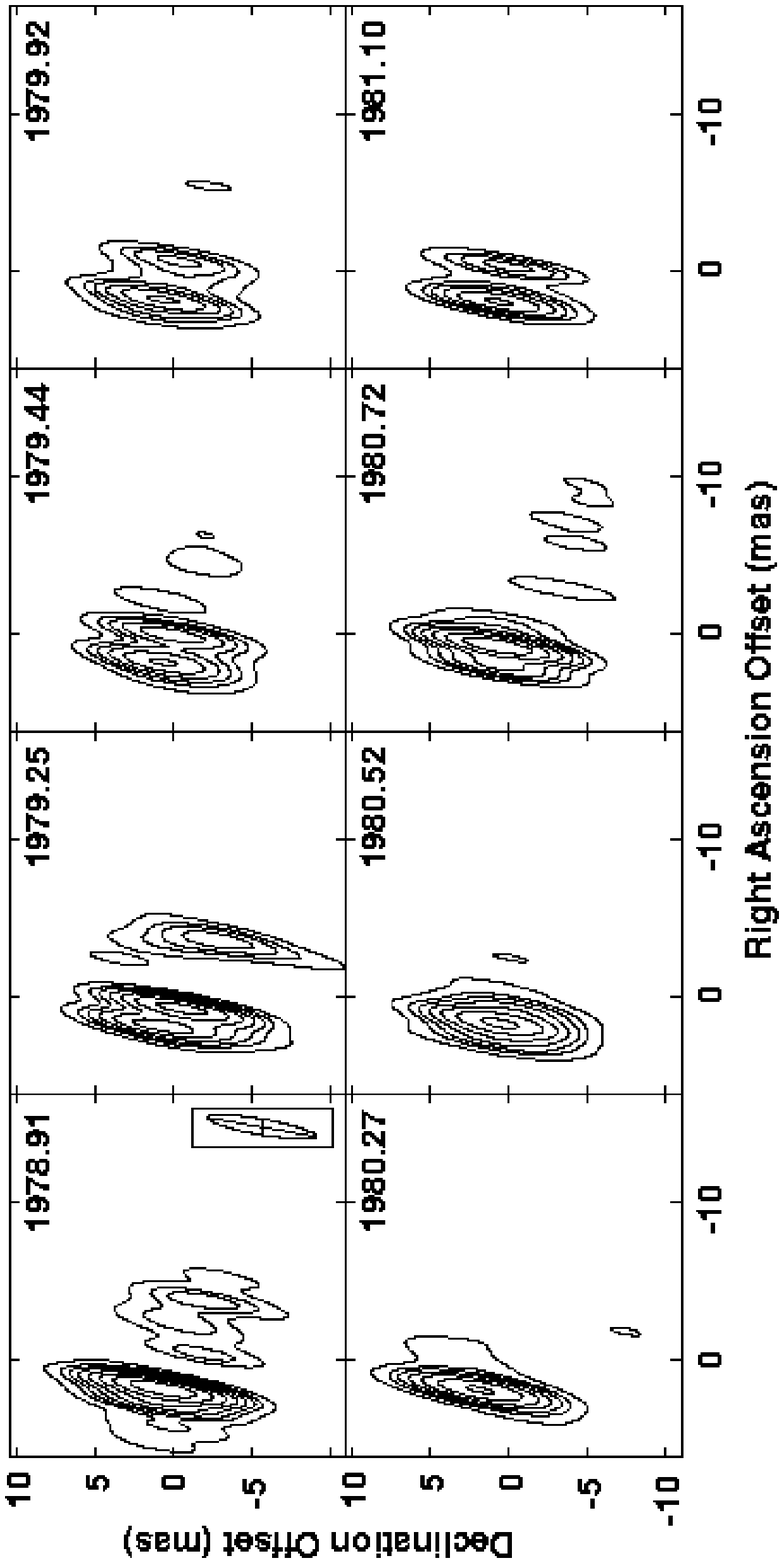}
\caption{The images from the 3C~120 Network VLBI monitoring
program between 1978.91 and 1981.10.  The images were made at either 5
or 10.7 GHz; see Table~\ref{obs6} for details.  All are images made
from CLEAN components convolved with a elliptical Gaussian beam 7.0 by
1.0 mas, elongated along position angle $-10\arcdeg$.  The contour
levels are 0.1, 0.2, 0.28, 0.4, 0.57, 0.8, 1.13, and 1.60 \jb.
The peak flux densities are tabulated in Table~\ref{obs6}.
\label{montage1}}
\end{figure}

\begin{figure}
\epsscale{0.55}
\plotone{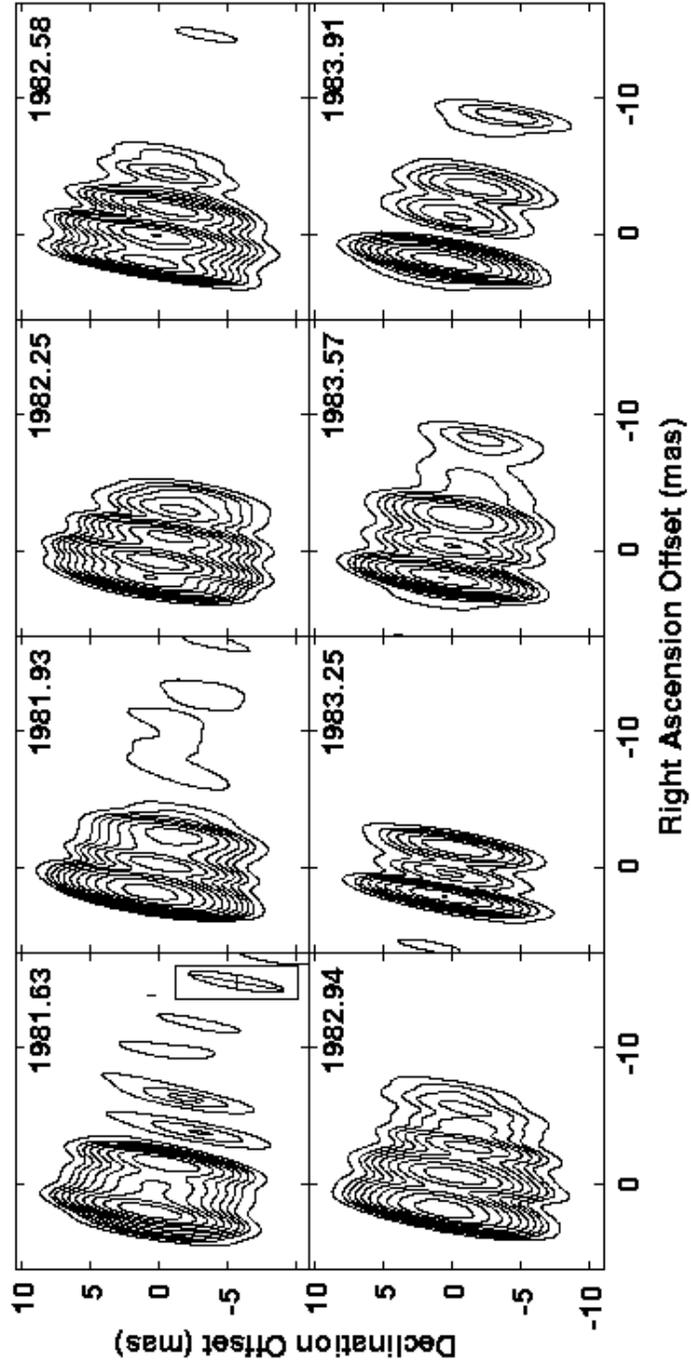}
\caption{The images from the 3C~120 Network VLBI monitoring
program between 1981.63 and 1983.91.  The images were made at 5 GHz.
All are images made from CLEAN components convolved with a elliptical
Gaussian beam 7.0 by 1.0 mas, elongated along position angle
$-10\arcdeg$.  The contour levels are 20, 40, 57, 80, 113, 160, 226,
320, 453, 640, and 905 m\jb.  The peak flux densities are tabulated in 
Table~\ref{obs6}.
\label{montage2}}
\end{figure}

\begin{figure}
\epsscale{0.75}
\plotone{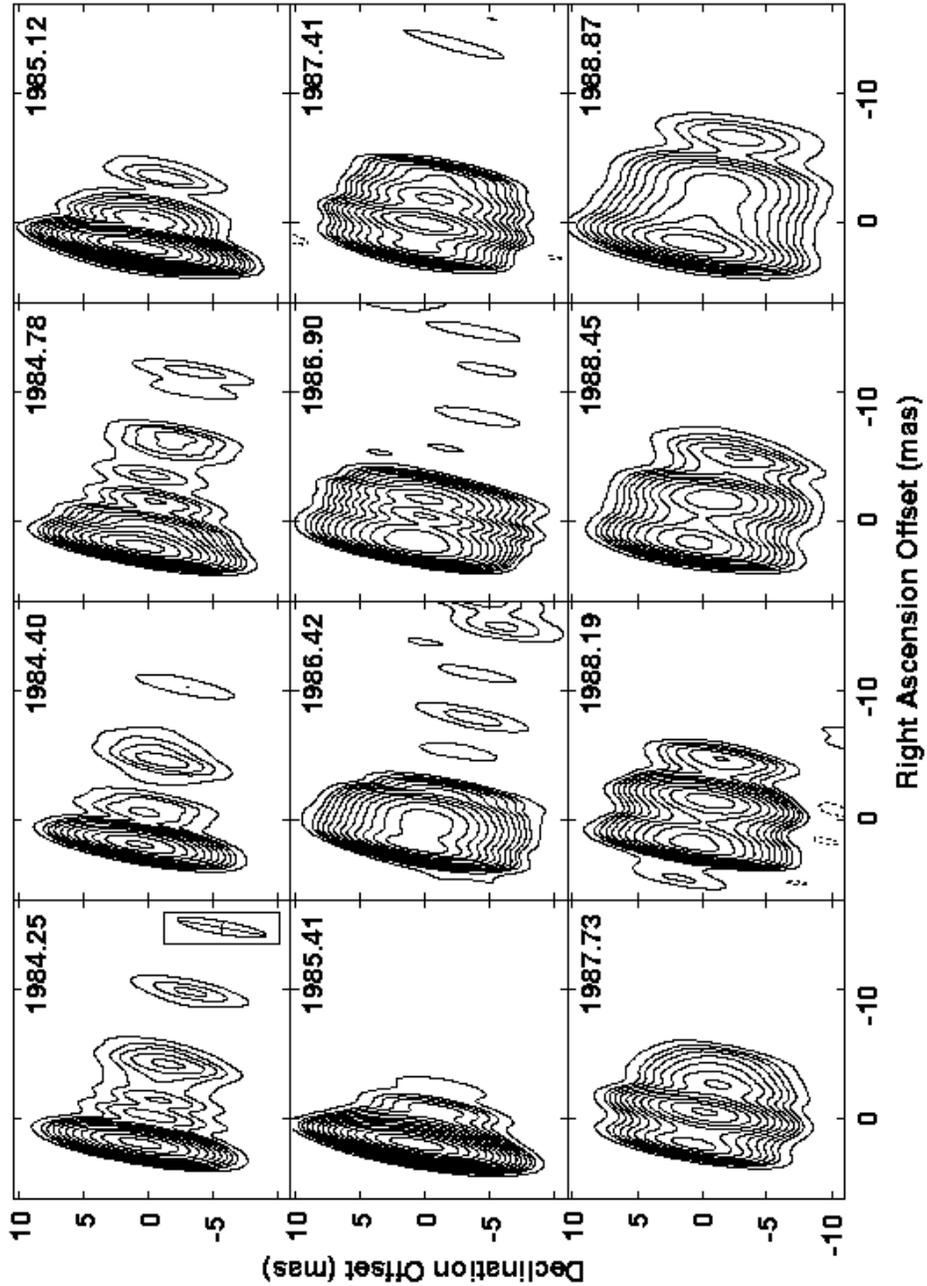}
\figcaption{The images from the 3C~120 Network VLBI monitoring program
between 1984.25 and 1988.87.  The images were made at 5 GHz.  The
contour levels are -10, 10, 20, 28, 40, 57, 80, 113, 160, 226, 320,
453, 640, 905, 1280, and 1810 m\jb.  The peak flux densities are
tabulated in Table~\ref{obs6}.  All of the images are made from CLEAN
components convolved with a elliptical Gaussian beam 7.0 by 1.0 mas,
elongated along position angle $-10\arcdeg$.  Half have the CLEAN
residuals restored to the image (1984.78, 1986.42, 1987.41, 1987.73,
1988.45, and 1988.87), but with the deep CLEANs used, this does not
make much difference.  The average off-source noise level in the images
with residuals is 1.3 m\jb.  That noise is dominated by calibration
artifacts.
\label{montage3}}
\end{figure}

    The images, consisting of CLEAN components convolved with a common
restoring beam of $ 7.0 \times 1.0 $ mas in position angle $-10\arcdeg$,
are shown in Figures~\ref{montage1} to \ref{montage3}.  The extremely
elongated beam is the result of the the low declination of the source
($5\arcdeg$) and fact that nearly all of the antennas available at the
time lay along a NE/SW line from Europe to California.  For most
epochs, the residuals are not included, but this does not make a
significant difference at the contour levels used.  Some of the more
recent images do include the residuals.  The quality of the images
improved over the years, as is reflected in the differing lowest
contour level displayed in each of the figures.

\placefigure{montage1}

\placefigure{montage2}

\placefigure{montage3}

    Figure~\ref{organ} shows the component motions.  A straight slice
was made in position angle $-105\arcdeg$ along the ridge line of the
jet in each of the images.  The parts of that slice that appear to be
reliable for identification of features were fit using between 2 and 6
Gaussian components.  In Figure~\ref{organ}, the amplitudes along each
slice are displayed as the width of the shaded structure centered at
the date of observation.  The fitted Gaussians were used to produce a
model for the amplitudes along the slice.  This model is shown as the
thin lines that appear to outline each of the shaded regions.  The
fact that the lines appear to outline the shading, rather than either
having shading extend outside the lines, or having white space inside
the lines, indicates that the Gaussian fits produce a good description
of the slice data.  Note that the lines only extend over the region
where structures are thought to be reasonably reliable, the region
over which the fits were made.  Especially in some of the earliest,
and least reliable epochs, the fits were not extended over all regions
that seem to show emission.  The fitted position of each component is
marked along with an error bar.  For the core, that error bar is the
maximum of the formal fit error for that component and a somewhat
arbitrary minimum error of 0.3 mas meant to account for alignment
uncertainties as discussed below.  For other features, the error is
the geometric sum of the core error and the formal fit errors for the
component position.  For the stronger features, the errors are
dominated by the assumed alignment uncertainty.  Thus the error bars
are meant reflect the full uncertainty of the position of each
component.

\placefigure{organ}

    The slices have been aligned by shifting each one so that the
position of the easternmost Gaussian, which is presumed to represent
the core, is set at zero offset.  Therefore the positions of other
components are relative to the core.  Some such alignment scheme is
required because, without use of phase calibration using a reference
source, there is no absolute position information available.  But no
alignment scheme will be perfect.  The apparent core position can be
influenced by emerging jet components while they are still too close
to be resolved from the core.  Such a component would shift the
apparent core toward the jet, which would make the apparent distance
to other jet features smaller.  The apparent core position can also be
influenced by variations in the length of the optically thick region
at the base of the jet.  This region is expected to be shorter at
higher frequencies so the distance to jet features is likely to be
higher when measured at 10.7 GHz rather than 5.0 GHz.  Examination of
Figure~\ref{organ} shows that the alignment would not be significantly
different if it were based on an assumption of linear motions of jet
components.  This suggests that the apparent core position does not
vary significantly.  Estimating by eye, the core position probably
does not vary by more than about $0.3$ mas, and that value is used as
the minimum core position error.

     Most of the components whose positions are shown in
Figure~\ref{organ} are marked as belonging to a specific feature seen
at multiple epochs and labeled with a letter.  Such components are
marked with closed circles.  Components marked with open circles in
Figure~\ref{organ} are not included in any motion analysis.  The
positions of all components belonging to each moving feature were fit,
in a least squares sense, for a feature angular speed and a start
date.  Straight lines indicating the fitted motion are drawn on the
figure.  The rates for each feature, along with the formal fit errors,
are displayed on the figure.  The fit results, including the start
dates, are also given in Table~\ref{speeds}.  The errors on the start
date are not symmetric.  Typically the uncertainty extends
significantly farther toward earlier than toward later times.  The
quoted errors were determined by exploring Chi-squared as a function
of the start date.  None of the error bars should be taken too
seriously.  They are correct under the assumptions that the motion is
linear, that the decomposition of the structure into Gaussian
components is appropriate, and that the images are reliable even in
the regions of relatively weak emission.  It is unclear how to
incorporate uncertainties related to these assumptions, but a
reasonable guess would be that they double the possible rate errors.

      It has been noted that, especially during the period between
1981.63 and 1984.40 when the observations occurred at very regular
intervals, there is a possible aliasing or ``stroboscopic'' effect.
The data could be explained in terms of features moving toward the
core or even by features moving outward at about 3 times the apparent
rate.  The position fits for such cases are not as good as for the
unaliased outward motion and the apparent expansion of the outer
envelope of observed emission is not as easily understood.  But such
cases are not excluded.  However at other times, especially late in the
project, the spacing of epochs was not so regular and it is reasonably
clear that such a stroboscopic effect is not happening.  Also the
assumed direction and rates are consistent with the 1.7 GHz results
presented below and with the high-resolution, frequent-sampling
results of \citet{G00}, both of which are much less subject to such
effects.

\placetable{speeds}

There is what appears to be a gap in the monitoring of features at 5
GHz at the beginning of 1986.  That is the result of a couple of
circumstances.  First, that time was preceded by a period of very low
activity in the source so there were no features along the jet with
any significant brightness.  Second, the core flared at that time,
starting new features, but also making it harder to see any more
distant features.  Finally, due to scheduling and processing problems,
there is a full year with no observations.  As a result of these
factors, the monitoring effectively started from scratch after 1986.

\subsection{Observations at 1.7 GHz}

\placetable{obs18}

     There have been 5 VLBI observations of 3C~120 at 1.7 GHz as
summarized in Table~\ref{obs18}.  Images from the first two have been
published in \citet{B88}, although they were remade with more modern
imaging methods, such as the use of robust weighting \citep{B95}, for
this paper.  All of these observations involved a large number of
antennas.  However the low declination of the source, combined with
the wide longitude range of the antennas, makes the coverage of the
(u,v) plane rather poor, especially for the longer baselines.  The
beams are not as seriously elongated as at 5 GHz because antennas off
the NE/SW line were recruited.  But other holes in the coverage
remain.  See the plots in \citet{B88} for examples. The first 3
observations used the Mark II VLBI system with 2 MHz bandwidth and 1
bit sampling.  The 1982.78 data were correlated on the Caltech 5
station correlator.  The 1984.26 data were correlated on the NRAO 3
station correlator as part of what has since become known as the
``World Radio Array'' set of observations of 4 sources.  The 1989.85
data were correlated on the Caltech Block II correlator.  The final
two epochs were recorded on the much wider bandwidth VLBA and Mark IV
recording systems.  The 1994.44 observation was made with just the 10
antennas of the VLBA \citep{N94} using a total of 32 MHz bandwidth and
1 bit sampling.  The 1997.71 data used a global array and both VLBA
and Mark~IV recording systems.  The total bandwidth was again 32 MHz,
but this time 2 bit samples were used.  Both the 1994.44 and 1997.71
observations were correlated on the VLBA correlator.  The improvement
of data quality with the wider bandwidth systems and the better (u,v)
coverage of the VLBA is readily apparent in the images.

\placefigure{big18}

\placefigure{medium18}

\placefigure{small18}

  The 1.7 GHz images are displayed on 3 different scales in
Figures~\ref{big18}, \ref{medium18}, and \ref{small18}.
Figure~\ref{big18} shows images from the two wide bandwidth
observations over the full region in which emission was detected.
That region extends to about 0\farcs5.  These images were made with
tapered data to give relatively low resolution to help draw out the
larger scale structure.  The beam was $18 \times 13$ mas at position
angle $3\arcdeg$.  These images show that the jet is a continuous
structure from the core region to at least a 0\farcs5.  In fact, it is
known to be a continuous structure from milliarcseconds to about
14\arcmin\ \citep{W87}.  The current images show much more clearly than
before that the jet has a smooth envelope; the edges follow smooth
curves.  However, there is considerable structure in the emission
internal to the jet, with significant side-to-side variations and
pronounced knots along the jet.

   Figure~\ref{medium18} shows all 5 epochs with a window that
includes everything out to 360 mas, including the big bulge at a bit
over 200 mas.  All images have been made using a Gaussian restoring
beam of $10 \times 4$ mas FWHM elongated in a position angle of
$-10\arcdeg$.  Note that the larger scale regions are much smoother in
the recent images than in the ones based on the older Mark~II data.
This is not a real effect in the source, but a reflection of the
behavior of the CLEAN deconvolution algorithm when dealing with a
combination of lower signal-to-noise ratio data, poorer (u,v) coverage
(especially missing short spacings), and probably higher closure
errors in the older data.  The closure errors are violations of the
assumption made in self-calibration that all calibration errors are
station dependent, affecting all baselines to a station equally.
The ridge of low level signal to the south of the core
in 1982.77 and the various peaks surrounding the core in 1989.85 are
artifacts resulting from closure errors.  Also, from the experience
gained while making the 1989.85 image, there appeared to be some
uniqueness problems in the region of the jet between about 20 and 75
mas, so subtle details in that area should not be taken too seriously.

  Figure~\ref{medium18} clearly shows that the bulge at about 200
mas is moving away from the core.  The exact motion is hard to
determine because the feature is large, without well defined compact
components identifiable over the whole time.  But estimates based
on the outer edge, the southernmost extension, and an eyeballed
estimate of the centroid, all constrain the total motion to be
between 20 and 32 mas for a rate of $1.7 \pm 0.4$ \masr, or
$3.5 \pm 0.8$ c for our assumed $H_0$.

  Figure~\ref{small18} shows the central regions of the same images
shown in Figure~\ref{medium18}.  For this figure, the vertical spacing
between the images is proportional to the time separation between the
observations.  This explains why the 1982.77 and 1984.26 images have
been cropped to allow them to be placed close together.  With this
vertical placement, any feature moving at constant velocity primarily
in the horizontal direction will move along a straight line whose
slope gives the speed. The alignment of the images along the jet
direction has been done by aligning the image peaks.  This is a
reasonable choice, although if the peak feature is resolved, there may
be a small shift from the optimum position.  This may be a concern for
1984.85 and 1997.70.  Any such offset should be less than about 2 mas.

  In Figure~\ref{small18}, dark vertical lines have been drawn at the
presumed core position and at the position of an apparently stationary
feature that will be discussed below.  Light shaded lines have been
drawn to highlight what appear to be moving features.  These lines are
hand drawn suggestions of how to link up features.  Clearly the
structure of these images would not lend itself well to the Gaussian
fitting procedures used at 5 GHz, especially in the regions more
distant from the core.  The same would probably be true at 5 GHz with
similar quality images that showed the structure at larger core
distances.  An additional difficulty is that the sampling interval was
sufficiently long during the 1984 to 1994 period to make association
of features a bit problematic.  Therefore the exact evolution of
features cannot be precisely specified.  But the lines are reasonable
suggestions.  All have slopes indicating feature speeds of between 2.5
and 3.1 \masr (5.1 and 6.3 c) with uncertainties of about 10\%.  These
speeds are similar to those of the faster 5 GHz features.  Given the
complexity of the structures and the difficulty of identifying exactly
how to track features, these speeds are all the same to within the
errors.

Note that two of the suggested features appear to pass through the
stationary feature at 81 mas, brightening and moving to the south side
of the jet in the process.  Also note that the ridge line of the jet
follows something of a twisted path that varies less between epochs
than do the individual features.  In particular, the bright region is
along the northern side of the jet in the 60-70 mas region and farther
out between 100 and 150 mas.  The stationary feature at about 81 mas
is brightest toward the southern side of the jet.  These sorts of
structures will be discussed below in terms of a possible helical
path.

\section{Discussion}

The images presented in this study, along with the higher resolution,
high frequency images of \citet{G98}, \citet{G99}, and \citet{G00}
make it clear that the 3C~120 jet is a continuous structure with
brightness fluctuations that move down the jet at apparent
superluminal speeds.  The fact that the jet is a continuous structure
is especially clear in the later (better) 1.7 GHz images which have a
fairly smooth outer envelope despite strong fluctuations of brightness
along the jet.  Identifiable features in the brightness fluctuations
can be followed for long distances as they move along the jet.
Feature B was followed at 5 GHz from 1 to 15 mas over 5 years.
Feature L7, seen at 1.7 GHz is likely to be a blending of several of
the 5 GHz features including B and D and it is tracked to 50 mas from
the core.  Thus, whatever the brightness fluctuations are, they
persist over large fractional core distances, suggesting that they
mark a change in some fairly important jet parameter.  A likely
possibility is that they reflect changes in the amount of material
moving down the jet.  The enhanced brightness could be simply the
result of increased density of emitting material.  But it is also
possible that a density change sets up conditions for shocks, which
might be what is actually observed.

The big bulge seen moving down the jet between 200 and 250 mas at 1.7
GHz may represent an extreme case of a density enhancement.  This
feature appears almost like a lobe.  It is seen best in
Figure~\ref{big18} where it clearly has a well defined outer edge and
it has clearly pushed out the envelope of the jet to the south.  This
is reminiscent of the 15 mas scale jet in 3C~84 \citep[and references
therein]{W94} where a similar lobe-like structure can be traced back
to a large flux density increase in about 1959.  In both cases, the
``lobe'' is penetrating into a preexisting jet structure.  The changes
in jet characteristics are sufficiently pronounced that the appearance
winds up similar to a large scale lobe interacting with an external
medium.  Note that the lobe speed of 1.7 \masr\ is somewhat less than
the speeds of most of the components measured on smaller scales.  This
situation is also seen in 3C~84, although that source is not
superluminal.  This is consistent with the ``lobe'' structure being a
region where material is piling up as it interacts with other, perhaps
slower, material in the older jet.  If this feature does represent a
period of past increased activity, that occurred roughly 100 years ago.

The 5 GHz observations suggest that new superluminal components are
emitted from the core of 3C~120 at intervals of under a year.  In
fact, this interval is mainly determined by a combination of the speed
of the jet components and the angular resolution of the observations.
At the observed angular rates of motion, features separated by
significantly less than a year would not be resolved into separate
components in our data.  The higher resolution observations of
\citet{G98} clearly show that there is complex structure on scales
below our resolution.  The birth of the brighter 5 GHz features is
indicated by a brightening of the core at around the time of zero
separation.  This is likely also true for the weaker features, but the
effect is disguised in the general variability of the core.  After an
initial brightening as they leave the core, the 5 GHz features
decrease in brightness as they move down the jet.  That decrease is
monotonic in these data.  The main exceptions seen in
Figure~\ref{organ} are the result of what is likely to be the poor
amplitude calibration in 1983.25 noted earlier.  However note that the
1.7 GHz data presented here and the high frequency data of \citet{G00}
show that a monotonic flux decrease is not a general property of
components in the source.

The core is often, but by no means always, the brightest feature in
the source.  Some of the brightest moving features remain brighter
than the core for up to about a year.  For this reason, any attempt to
do accurate geometric work such as geodesy or astrometry with this
source must take structure into account in order to obtain results
accurate to better than about 3 mas.

The speeds of features measured from the images presented here suggest
that there is some variation in speed between components, but that
variation is under a factor of two.  At 5 GHz, speeds of between 1.5
and 3.5 \masr\ are measured, but that range shrinks to 2.2 to 3.2 if
only features seen in more than 3 images are included.  The measured
speeds are based partly on the assumption that the brightness
fluctuations can be modeled as Gaussian components.  That is clearly
simplistic, and much blending of structures is involved.  While the
possibility that all features have the same speed is strongly excluded
by the formal errors of the fits, it is not absolutely excluded when
the reliability of the images and the difficulty of identifying some
of the features at larger core distances are taken into consideration.
However, note that Feature~C, which is formally slower than Feature~D,
seems to disappear when Feature~D catches up, supporting the idea that
the speeds really do vary.  Also the comparatively low speed of
Feature~J is reasonably convincing.  In the 1.7 GHz images, all of the
moving features less than 180 mas from the core have speeds similar to
those of the 5 GHz features and to the speeds seen at 22 GHz close to
the core in the movie of \citet{G00}.  There is a significant change
at the 200 mas bulge, discussed above, which is clearly moving
more slowly than features closer to the core.  There is no clear
evidence at either 1.7 or 5 GHz for a change of speed of an individual
component.

The 3C~120 jet undergoes a significant bend on small scales.  The high
resolution images of \citet{G98} and \citet{G99} show that it is first
seen at submilliarcsecond scales to have a position angle of about
$-120\arcdeg$.  Our 1.7 GHz images show an angle of roughly
$-100\arcdeg$ out to about half an arcsecond.  The high resolution
images suggest that most of the curvature between these values occurs
between about 2 and 5 mas.  However our 5 and 1.7 GHz observations
suggest that a position angle of near $-105\arcdeg$ is maintained as
far as about 30 mas, at least in the more recent data.  Some of the
apparent curvature between 2 and 5 mas may be the result of internal
side-to-side structure such as those at larger scales discussed below.
Higher dynamic range, high resolution observations will clarify this.

It would be interesting to know how constant the position angle has
been, especially on the small scales near the core.  But unfortunately
the very poor north-south resolution of the 5 GHz images makes it
difficult to obtain accurate position angle measurements.  The images
of Figures~\ref{montage1}, \ref{montage2}, and \ref{montage3} do show
that the position angle did not change much between 1979 and 1988.
All images show position angles between the main features of between
about $-102$ and $-108\arcdeg$.  A 1999 image at 5 GHz, currently
being processed (Benson, private communication), and the high
frequency observations cited above, show a similar position angle
except on scales of less than about 2 or 3 mas.  On the smallest
scales, the position angle is $-120\arcdeg$ or more negative.  This
could indicate that a real change in the position angle of the jet on
the smallest scales has occurred.  But it could simply show that the
newer data, with improved north-south resolution, is better able to
determine the position angle on these scales.

The evolution of components at 1.7 GHz is somewhat different from that
seen at 5 GHz.  The general decrease of brightness along the jet is
apparent.  But individual features can vary both up and down in
brightness.  In fact, there may be a pattern to where such variations
occur, although the combination of the limited quality of the early
epoch images and the poor time sampling make full delineation of such
patterns problematic.  The most probable case is the bright feature at
about 81 mas that is there at some level in all 5 images.  It appears
that both features L4 and L5 pass through this feature.  About 10 mas
toward the core from this feature is a region that is relatively weak
in all of the images.  The bright feature is biased toward the
southern side of the jet, while the weaker regions both closer to the
core and farther out are to the north.  This appears to be a
stationary pattern superimposed on the moving jet.  An eyeball
estimate of the upper limit to the pattern speed is about 0.55 mas/yr
or about $1.1 c$, much slower than the speed of the knots.  There are
much less convincing hints of a similar pattern near 45 mas from the
core.  Such stationary patterns could result from a curved trajectory
or from standing shocks, perhaps related to interaction with an
external medium.

\section{Implications of a Helical Pattern}

\begin{figure}
\epsscale{1.0}
\plotone{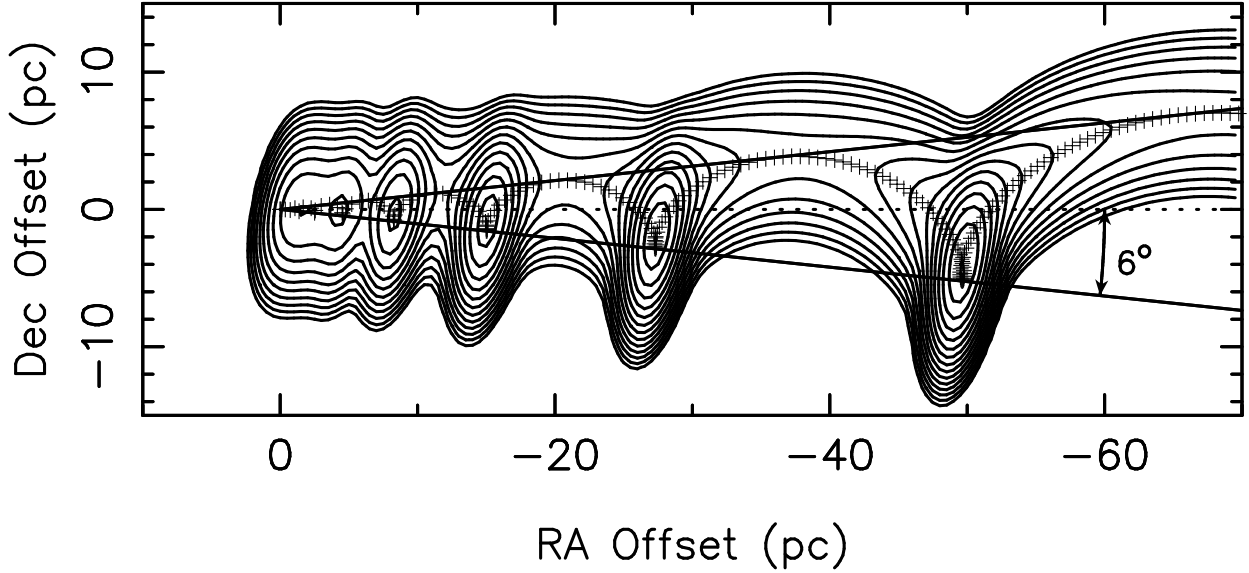}
\caption{A contour image, with $\sqrt{2}$ levels, of the
predicted structure of a simple helical jet.  The crosses mark the
sky-plane positions of test emitters at the time that the observed
wave-front passed them.  The test emitters follow a fixed helical path
on the surface of a cone and are of constant brightness in their rest
frames.  The emission from the model points has been convolved with
the 12.5 by 4.0 mas beam used in the 1.7 GHz images, scaled by 0.6 pc
mas$^{-1}$.  The solid lines outline the cone while the dotted line
indicates the cone axis.  The model is the one described in the text
without any offset between the brightness ridge and the path of the
particles.  Other models with the particles following narrower spirals
at higher angles to the line-of-sight, but with the brightest emission
shifted toward the outside of the jet, produce very similar looking
images.  \label{helix}}
\end{figure}

The patterns described at the end of the last section are similar to
those that would be expected if the brightest part of the jet followed
a helical path on a cone.  The brightest regions would be on the side
where the path approaches closer to the line-of-sight.  The brightness
would be enhanced both by foreshortening, if the jet is optically
thin, and by relativistic beaming, which increases at smaller
line-of-sight angles.  On the other side of the cone, the path would
both be stretched out by projection and the brightness would be
decreased because of reduced beaming.  Figure~\ref{helix}, to be
discussed in more detail later, shows the predicted appearance of a
smooth jet following such a helical path.  The region of the 3C~120
jet in the vicinity of the 81 mas (49 pc) feature looks much like the
model, except that the jet is not smooth and components move along the
path passing through the stationary pattern.  The helix would be
within the overall envelope of the jet since it is indicated by the
line of brightest emission, not by the rather smooth outer limits to
the jet sides.  

Helical patterns are expected to occur in jets as the result of
Kelvin-Helmholtz instability \citep[and references therein]{H87,H00},
possibly driven by small perturbations near the central engine, so it
would not be surprising to see a helical pattern in 3C~120.  In this
section we will discuss the implications of the observation of a
helical pattern.  But it should be kept in mind that the present data
do not exclude other explanations for the structures seen.  For
example, shocks are often invoked in discussions of the nature of
superluminal features and other brightness fluctuations in jets.  It
is possible that the structures seen are the result of a combination
of moving and stationary shocks.  Monitoring at 1.7 GHz is continuing
in an effort to improve the observational constraints.

\citet{S95} have explored the implications of the possible helical
paths of superluminal components in 3C~345.  They consider the the
helical trajectories implied by 4 different cases of conserving 3 of 4
possible quantities; kinetic energy, angular momentum, linear momentum
in the jet direction, and jet opening angle.  The 3C~345 trajectories
open out toward straight lines which are better matched by their
Case~3 in which the unconserved quantity is linear momentum along the
jet axis.  However the 3C~120 morphology, which shows suggestions of
multiple cycles of the helix, looks more like their Case~2, which
allows the angular momentum to vary.  They note that this case is
equivalent to that of the hydrodynamic isothermal helical model of
\citet{H87}.

The possible observation of a slowly moving, long wavelength, helical
pattern has implications for the jet orientation and for the nature of
the medium through which the jet propagates.  To explore these
implications, we first consider a simple morphological model which can
be used to deduce some deprojected properties of the jet under the
assumption that the helical pattern actually traces the path taken by
jet material.  We then consider how such a model relates to the flow
through a helical pattern deduced from the time dependent linearized
fluid equations.  The theory suggests that an observed helical pattern
may imply larger transverse motion than is experienced by individual
fluid elements, thereby relaxing the constraints on the geometry.
Finally we consider the implications of a slow pattern speed.  A
summary of the models considered and their implications is given in
Table~\ref{models}.

\placetable{models}

\subsection{A Simple Model}

For the simple model, we assume that the jet follows a helix that lies
along the surface of a cone and has a wavelength that is a constant
multiple of the cone radius, that the pattern is fixed, that the flow
follows the pattern, and that the jet is optically thin.  This
corresponds to Case~2 of \citet{S95}.  The observed cone opening angle
and helix wavelength constrain the true opening angle and wavelength
as a function of the angle of the cone axis (hereafter called the jet
axis) to the line-of-sight, $\theta$.  The observed superluminal
motion constrains the jet speed in the direction of the jet axis,
since it is an average speed.  But the jet material has velocity
components in other directions in order to follow the helix, so the
required $\gamma$ is somewhat higher than the usual $\gamma \approx
\beta_{obs}=v_{obs}/c$ ~of a straight jet and is constrained by the
geometry as a function of $\theta$.  With these constraints, the
minimum and maximum angles to the line-of-sight of the flow along the
pattern, and hence the beaming factors and projection effects, can be
determined as a function of $\theta$.  The implied brightness
difference between the brightest and dimmest parts of the pattern is a
strong function of $\theta$ in this model and so can be compared with
the images to constrain that angle.

To match 3C~120 with calculations of the sort described in the next
paragraph, we used an apparent (projected) half opening angle
of the cone of $6\arcdeg$, a helix wavelength such that a full cycle
occurs between the possible 45 mas feature and the 81 mas feature (27
and 49 pc at the assumed distance), an apparent average speed of 5.7
c, and a ratio of brightnesses of 20.  We assumed a typical optically
thin jet spectral index of $\alpha=-0.7$ ($S\propto \nu^{\alpha}$).  A
reasonable match to all of these constraints is given if the cone has
an intrinsic half opening angle of $0.16\arcdeg$, is at $1.5\arcdeg$
to the line-of-sight, has a helix wavelength of $\lambda = 120 R$ (R
is the jet radius) and has $\gamma = 9$.  One fairly general property
of a helix with the assumed projected wavelength and opening angle is
that the path swings almost directly into the line-of-sight on one
side.

Figure~\ref{helix} shows the appearance of a simplified jet with the
above properties.  The ``jet'' consists of a continuous stream of
unresolved, optically thin emitters following a fixed helical path 
on the surface of a cone at
constant speed.  The emitters are started along the jet at equal
intervals and are of constant brightness in their rest frames.  The
position of each emitter, at the time the observed wavefront passes
it, is calculated along with its beamed brightness.  The crosses in
the figure show the ``observed'' positions of the emitters on the
plane of the sky.  No attempt has been made to account for brightness
evolution as the emitters move down the jet (dimming expected) or to
model the brightness variations along the jet that are seen as
components in the real data.  The beamed flux density from each
emitter is convolved with an observing beam, equivalent to that used
for the 1.7 GHz images (with our assumed $H_0$), and added to the
model image.  The brightness enhancements where the jet swings into
the line-of-sight on the southern side of the jet are obvious as are the
dimmer looping structures along the northern side of the jet between
the bright regions.

The modeled jet has a high degree of collimation with a pattern that
only slightly deviates from longitudinal, but is both broadened and
foreshortened by the projection effects of being very close to the
line-of-sight.  It is considerably closer to the line-of-sight than is
required by the usual model for superluminal motion in a straight jet.
In the straight jet model, $\gamma \approx \beta_{obs}$ and $\theta$
is assumed to be about $1/\gamma$, although a fairly wide range of
angles is possible.  For 3C~120, a straight jet would have $\gamma
\approx 5.7$ and $\theta \approx 10\arcdeg$ with values up to near
$20\arcdeg$ possible.  But for larger angles such as these, a helical
jet matching the observed speeds and opening angle would have much too
high a brightness contrast between the peaks and regions between the
peaks.

There are at least two ways that the brightness contrast might be
reduced, thus allowing higher angles to the line-of-sight and less
severe collimation.  It is possible that not all of the emitting
material participates in the helical path, as is suggested by the
relatively smooth outer jet boundary.  Such emission could dominate
between the peaks, reducing the observed brightness ratio.  However in
that case, it seems likely that the brightness would be distributed
more evenly across the jet between the peaks, contrary to the
observations.  The other possibility is that the observed opening
angle of the jet does not reflect the actual deviations of the
material flow from the direction of the jet axis.  The high pressure,
and hence brightest, region may swing more widely than the actual
particles as the jet flows through the helix.  This would reduce the
differential beaming effects and smooth the appearance of the jet.  As
noted below, physical models suggest that this will be the case.

\subsection{Physical Theory; Decoupling the Pattern and the Material Paths}
\label{ptheory}

Studies involving both physical theory and numerical simulation show
that helical patterns are likely to form in jets, including
relativistic jets.  \citet{H87,H00} has shown that helical patterns
are an expected result in many circumstances, can induce significant
transverse motion without the development of shocks, and can produce
asymmetries in Doppler boosted emission if jets are observed at
line-of-sight angles on the order of the beaming angle.  In
particular, the theory has been compared to observations, especially
of M~87, which shows helical patterns \citep{O89}.  But one difficulty
is that the theoretical helical patterns, and those observed in M~87,
are much more tightly wound than what we see in 3C~120.  The fastest
growing unstable modes in these calculations typically have
wavelengths of only a few jet radii.  These calculations generally
assume a hot jet in a hot medium (like a cocoon).  But the wavelength
of the fastest growing structures increases if the jet sound speed is
lower (higher internal Mach number) and/or the Lorentz factor is
higher, provided that the ambient (external) sound speed is much less
than the jet sound speed.  This suggests that the 3C~120 jet
resides in a relatively cool medium, and not in a relativistically hot
cocoon or lobe.

The linearized time dependent fluid equations give the distribution of
pressure and fluid velocity field within the jet as fluid moves through
a helical pattern.  Typically the highest pressures are near the
surface of the jet at an azimuthal position that rotates with the phase
of the pattern.  The actual jet material need not follow the path of
the region of highest pressure.  In general the flow lines will follow
a smaller diameter helical path than the region of highest pressure.
The brightest radio emission is expected to be generated in the region
of highest pressure so the observations will trace the path of that
region.  But one cannot use that path to calculate the relativistic
beaming effects because they depend on the vector velocity of the
emitting material.  In principle, the reduced helical path followed by
the material allows for lower brightness variations than are calculated
assuming that the apparent path is the material path.  A general result
of calculations from the fluid equations is that, as one moves to shorter 
pattern wavelengths, the transverse motions become smaller so the
suppression of beaming-induced brightness contrasts becomes larger.  

We have modified the simple model described earlier to allow for
material following tighter helices than the apparent path of the
brightest emission.  With this modified scenario, it is possible to
find parameters that predict an appearance and apparent velocity that
are very similar to the predictions of the model used above to fit the
3C~120 data, but with much wider angles between the average jet axis
and the line-of-sight.  A fit can be obtained with essentially any
angle to the line-of-sight up to at least $16\arcdeg$.  As that angle
increases, the required ratio of the radius of the apparent helical
path described by the brightest emission to the effective radius of
the helical path followed by the material increases.  For $10\arcdeg$
to the line-of-sight, the apparent helical path must be about 2.2
times the effective radius of the helical path followed by the
material.  For $16\arcdeg$, the ratio would be about 4.6.  

As the angle of the jet to the line-of-sight increases, the helical
wavelength required to approximate that seen in projection in the
3C~120 images decreases.  To compare with the theory, we need to
relate the helical wavelength to the jet radius, and it is no longer
obvious how that radius relates to features in the images.  But
assuming that the path of brightest emission marks the edge of the jet,
the wavelengths for the models at $10\arcdeg$ and $16\arcdeg$ to the
line-of-sight have wavelengths of 33 and 21 times the radius.  This is
much shorter than that implied by the original model with its very
small angle to the line-of-sight, but is still large compared to the
situation analyzed for M87.  Ultimately, this sort of analysis should
be done by including the radio emission, along with beaming effects,
in the physical theory or numerical models.

\subsection{Moving Patterns}

One difference between the simple morphological model and a helically
twisted jet is that the helical pattern will be moving.  Motion of a
large scale helical pattern can be computed from the linearized time
dependent fluid equations, e.g., \citet{H00}.  While in general the
pattern speed is a function of the wavelength, at very long wavelengths
relative to the most Kelvin-Helmholtz unstable helical wavelength (see
eq. [7b] in Hardee 2000), the pattern speed may be found from the real
part of (eq. [5a] in Hardee 2000)
$$
\omega /ku\approx \frac{\gamma ^2\eta _{rel}}{1+\gamma ^2\eta
_{rel}}\pm i\frac{\gamma \eta _{rel}^{1/2}}{1+\gamma ^2\eta _{rel}}~.
$$
Here $\gamma$ is the Lorentz factor of the underlying flow with speed $u$ and
$\eta_{rel}\equiv (a_{ex}/a_{jt})^2$, where $a_{ex}$ and $a_{jt}$ are
the sound speeds external to and inside the jet respectively.
The sound speed is 
$a\equiv [ \Gamma P_0 / (\rho _0+[\Gamma /(\Gamma-1
)]P_0/c^2) ] ^{1/2}$, 
where $4/3\leq \Gamma \leq 5/3$ is the adiabatic index.  The density,
$\rho_0$ and pressure, $P_0$, are measured in the proper fluid frame,
and since pressure balance has been assumed $\eta _{rel}$ is
effectively an enthalpy ratio.  At the most unstable wavelength
$$
\lambda \approx 3.75 \frac{\gamma
[M_{jt}{}^2-u^2/c^2]^{1/2}M_{ex}}{[M_{ex}{}^2-u^2/c^2]^{1/2}+\gamma
[M_{jt}{}^2-u^2/c^2]^{1/2}}R~,
$$
the pattern speed is approximately
given by (eq. [7c] in Hardee 2000)
$$
v_w\approx \frac{\gamma
[M_{jt}{}^2-u^2/c^2]^{1/2}}{[M_{ex}{}^2-u^2/c^2]^{1/2}+\gamma
[M_{jt}{}^2-u^2/c^2]^{1/2}}u~, 
$$
where $M \equiv u/a$. In general, we expect that any large scale
helical twist and observed helical pattern to have a pattern speed that
lies between the values at long wavelength and at the most unstable
wavelength.  In the absence of a low frequency precession at the
central engine, the most unstable wavelength is the one most likely to
grow and be seen.  Helical wavelengths shorter than the most unstable
wavelength will not move the whole jet bodily but can still produce a
helical pattern near to the jet surface.  In general, the shorter
helical wavelengths propagate with much higher pattern speeds.

A long stationary pattern wavelength such as we assume for 3C~120
suggests that $\eta_{rel}$ is small and that $M_{ex}$ is large, as
this sets an upper limit to any helical pattern speed to be
significantly less than $c$.  Because of projection effects, and a
speed that is too low to amplify the apparent speed as for
superluminal motion, such a pattern would appear to be stationary in
our data, consistent with our results.  Thus, a cool and not a
relativistically hot external cocoon or lobe medium is implied for a
long wavelength pattern that is relatively slowly moving.  This is the
same conclusion that was reached in section~\ref{ptheory} from the
desire to have the most unstable modes be of long wavelength to match
the observations.

While the helical pattern is not moving superluminally, the apparent
helical wavelength in the observer frame would be affected by any
motion.  The observed radiation travels nearly parallel to the jet so,
in the time it takes to go between points where the pattern is at the
same azimuth with respect to the jet axis, the pattern has moved.  The
apparent observed pattern is stretched as 
$$ 
\lambda^{obs}=(\frac{1}{1 - \beta_w cos\theta})\lambda 
$$ 
where $\beta_w=v_w/c$.  For example, a helix with a wavelength of
$80R$ in the pattern frame moving with a pattern speed $c/3$ at an
angle of $3^{\circ}$ to the line-of-sight has an observed wavelength
of about $120R$.  A second effect of a moving helix is that fluid does
not need to move as rapidly in the transverse direction to circulate
through the helix as the fluid does for a ``stationary'' helix. The
two effects will modify the Doppler boosting somewhat, although the
two effects will tend to cancel each other as the stationary helical
path has a longer intrinsic wavelength than the moving helical path.
To investigate these effects we have computed, from the linearized
time dependent fluid equations, the velocity components of material
moving through a $120R$ nearly stationary helix (pattern speed c/100)
and for a $80R$ moving helix (pattern speed c/3), both with transverse
helical displacement of the jet by 1/2 the jet radius.  There is very
little difference in the transverse velocity components for the two
cases. At least approximately (the flow angle is not exactly the same
over the jet cross section), the flow angle with respect to the jet
axis is about the same for the stationary helix and for the moving
helices.  Thus, we expect calculations for a stationary $120R$ helix
to be only slightly modified if a helical pattern speed is considered.
Similar effects would be expected for the shorter wavelengths allowed
by the decoupling of the pattern and material paths.  One goal of
extended monitoring observations is to provide much improved
constraints on the pattern speed.

Additional computations help constrain the angle of the jet to the
line-of-sight.  As noted earlier, the limited brightness contrasts
along the pattern cannot be used to do this because transverse motions
of the material may be considerably smaller than implied by the
observed pattern.  But the limits on the observed pattern speed can
provide constraints.  Larger line-of-sight angles require deprojected
helical pattern wavelengths that are shorter than the fastest growing
helical wavelength.  While we cannot completely rule out shorter
somewhat more rapidly moving helical patterns, preliminary
computations suggest that the observed pattern wavelength must be
longer than the most unstable wavelength if the observed pattern speed
is to appear stationary in the data.  Given the pattern speed
appropriate to the most unstable wavelength, we find that the shortest
deprojected wavelength consistent with the measured upper limit
to the pattern speed is given by
$$ \lambda \approx 3.75 \frac{\gamma
[M_{jt}{}^2-u^2/c^2]^{1/2}M_{ex}}{[M_{ex}{}^2-u^2/c^2]^{1/2}}R~. 
$$
In the above expression we have assumed that $cos \theta \approx 1$ and
that $(1 - u/c) << 1$.  Models involving line-of-sight angles larger
than about $10\arcdeg$ require a deprojected observed helical
wavelength shorter than this limit and thus would show too high
a pattern speed.  Note that these arguments imply that the tightly
wound helix seen in M87 would have a high pattern speed.

\subsection{Helical and Other Patterns Elsewhere in 3C~120}

Patterns reminiscent of a helix are seen on other scales in 3C~120.
On scales between about 1 and 15 arcseconds, there are several cycles
of an oscillating pattern that could be a helix \citep{W97}, again
with a long wavelength.  In fact the 4\arcsec\ knot, whose motions
were the object of study of \citet{W97}, could, from morphological
grounds, be another feature like the one at 81 mas.  This reinforces
the idea, mentioned in the introduction, that the lack of motion in
this feature does not indicate that the jet material has slowed.
\citet{G98} suggest that the structures observed at 22 and 43 GHz in
3C~120 at about 3 to 5 mas from the core are the result of a helical
path on a much smaller scale than what we see at 1.7 GHz.  As with our
results, the wavelength is long, even in projection, relative to the
jet radius.  But, more extensive monitoring has suggested another
explanation related to interaction with a cloud intermediate in mass
between narrow and broad line clouds \citep{G00}.  This provides
a cautionary note against over interpreting the current data and shows
the need for continued monitoring to clarify the situation.

The 3C~120 jet shows a number of pronounced curves that are probably
not part of a helical pattern.  The first is in the region of a few
mas where the jet bends from a position angle of about $-120\arcdeg$
\citep{G98} to the approximately $-100\arcdeg$ seen out to a few
arcseconds.  At about 6 arcseconds, the jet begins a long gentle
curve that eventually takes it through an angle of at least
$120\arcdeg$ over several arcminutes on the sky before the structure
becomes difficult to interpret \citep{W87,W97}.  These curves are
probably real bends in the jet.  Because of projection at the small
angle to the line-of-sight, such curves need only be through a few
degrees.  They may be the result of interaction with an external
medium as suggested by the pronounced side-to-side asymmetry between 5
and 25 arcseconds seen in the image in \citet{W97}.

\section{Conclusions}

\begin{figure} 
\epsscale{1.0} 
\plotone{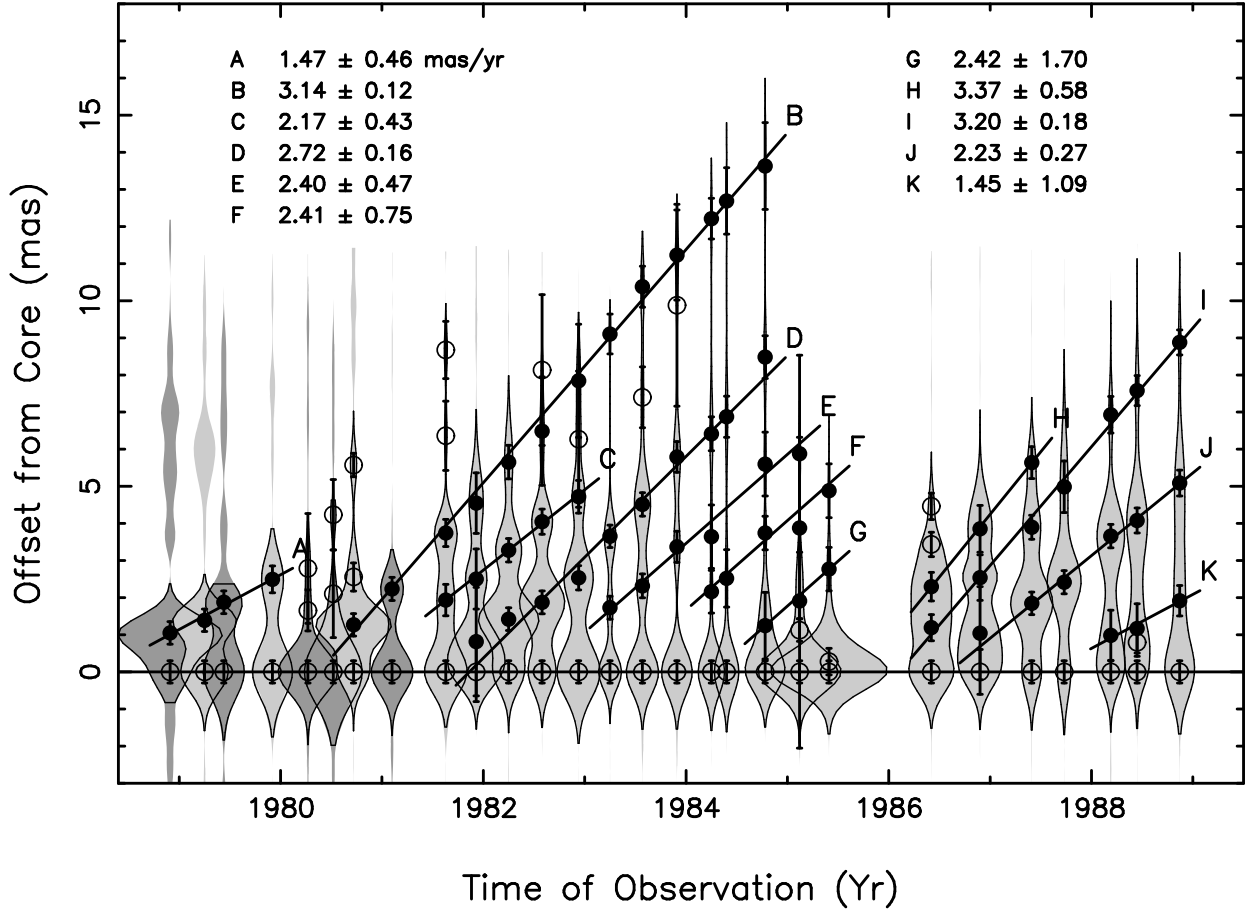} 
\caption{This figure
shows the motions of components seen at 5 and 10.7 GHz in 3C~120
between 1978 and 1988.  For each image, there is a shaded profile
whose width represents the amplitude along a slice that runs along the
ridge line of the jet.  The 10.7 GHz images are indicated by darker
shading.  The thin lines that appear to outline the shading are the
results of Gaussian fits to the reliable parts of the slices.  The
profiles have been aligned on the eastern-most feature, presumed to
be the core.  The fitted positions, relative to the core, of other
features are shown by circles with error bars that reflect the
combined formal position errors for that feature and for the core,
along with an assumed alignment error.  Features with filled circles
are ones that are identifiable at multiple epochs and for which speeds
are measured with a least squares fit.  The components are labeled and
a straight line with the slope corresponding to the fitted velocity is
drawn through the points.  The velocity for each feature, with formal
errors, is written on the figure and listed in Table~\ref{speeds}.  
\label{organ}} \end{figure}

\begin{figure}
\epsscale{1.0}
\plotone{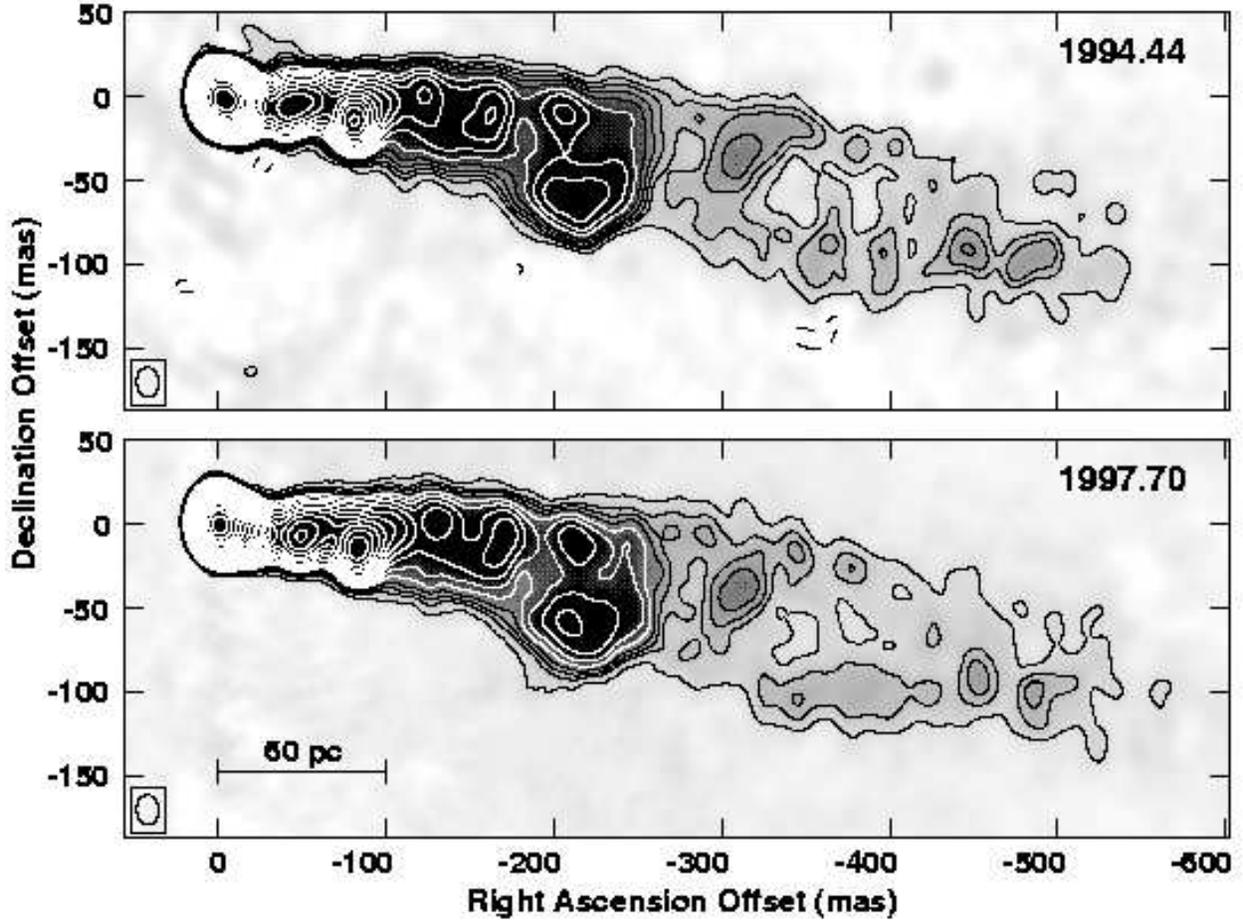}
\caption{Low resolution versions of the 1.7 GHz images
from 1994.44 and 1997.70.  These were the epochs that used wide band
recording systems and had the sensitivity to see the structures
between 270 and 540 mas from the core.  Both of these images have been
made with a taper that uses mainly the continental and shorter
baselines.  They were convolved with a common beam of 18 by 13
mas at position angle $3 \arcdeg$.  The contour levels are -1.4, -1.0,
-0.5, 0.5, 1.0, 1.4, 2.0, 2.8, 4.0 m\jb\  with the higher levels
increasing by factors of $\sqrt{2}$ from there.  The image peaks
are 1.42 and 1.83 \jb\ in 1994.44 and 1997.70 respectively.  The
off-source rms noise levels are 0.15 and 0.10 m\jb.
The scale bar is based on $H_0=75$ \kms Mpc$^{-1}$.  \label{big18}}
\end{figure}

\begin{figure}
\epsscale{0.70}
\plotone{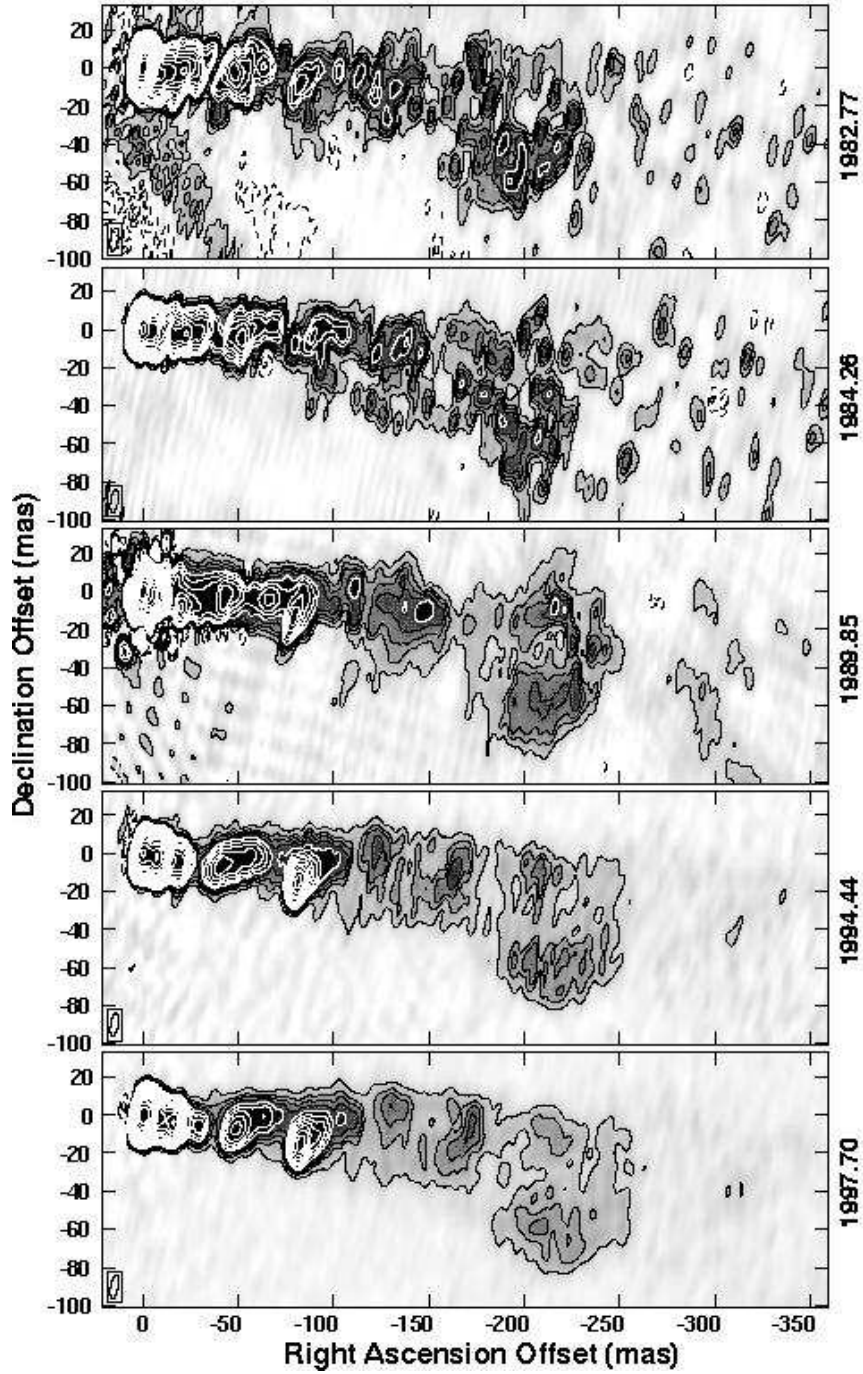}
\caption{The inner 360 mas of the 5 images available
at 1.7 GHz.  All have been convolved to a common beam of 12.5 by 4.0
mas elongated in position angle $-10 \arcdeg$.  The lower contour
levels are -2.1, -1.5, -0.75, 0.75, 1.5, 2.1, 3.0, 4.2, 6.0 m\jb\ with
the higher levels increasing by factors of $\sqrt{2}$ from there.  The
peak flux densities and noise levels can be found in Table~\ref{obs18}.
\label{medium18}}
\end{figure}

\begin{figure}
\epsscale{0.65}
\plotone{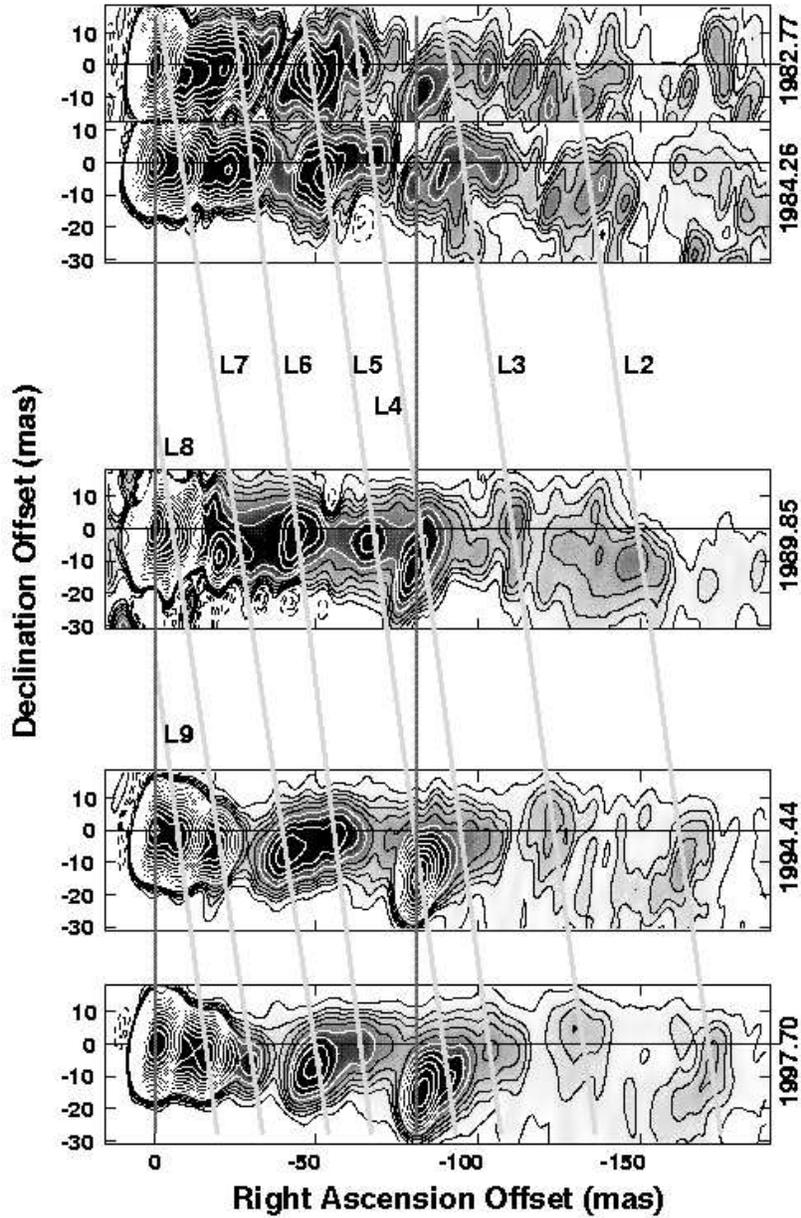}
\figcaption{The inner 190 mas of the 5 images available
at 1.7 GHz.  These are the same images with the same contour levels as
those in Figure~\ref{medium18}.  The images have been spaced
vertically on the page so that the separations between the cores is
proportional to the time intervals between the observations.  This
allows features moving with constant apparent speed to be marked with
straight lines.  The source appears to contain both moving and
stationary features.  A stationary feature and the core have been
marked with dark vertical lines.  The light lines are eyeball
connections of moving features.  All have slopes of between
2.5 and 3.1 \masr.
\label{small18}}
\end{figure}

   The main conclusions of this study are:

\begin{itemize}

\item The 3C~120 jet is a continuous structure with brightness
variations.  Identifiable features in the brightness structure move
down the jet at superluminal speeds.  We are able to identify new
features appearing at intervals of less than a year
(Figure~\ref{organ}, Table~\ref{speeds}), an interval that is set
mainly by a combination of the jet speed and the angular resolution of the
observations.  Finer scale structure is known to exist as shown by
higher frequency observations by other groups (c.f. G\'{o}mez et al. 2000)

\item The two best 1.7 GHz images show structure with large variations
of brightness along the jet, but a fairly smooth outer envelope of
the emitting region (Figure~\ref{big18}).

\item The speeds of the 5 GHz components in the region between 1 and
10 mas of the core appear to vary over a range that is well under a
factor of two (Table~\ref{speeds}).  The formal errors strongly
exclude the possibility that the speeds are all the same.  But because
of the limited image reliability and the difficulty in identifying
some of the features at larger core distances, that conclusion is not
as firm as the formal statistics would suggest.

\item On the larger scales sampled by the 1.7 GHz images, most of the
moving features have angular rates in the 2.5 to 3.1 \masr\ range, and
within the uncertainties, could all be the same
(Figure~\ref{small18}).  This is within the range of speeds seen at
higher resolution in our 5 GHz observations and in the 22 and 43 GHz
observations of \citet{G98}.  The fact that the speeds show the same
fairly narrow range between a few tenths of a parsec to about 120
parsecs in projection indicates that there are not large changes in
either the underlying speed or angle to the line-of-sight of the jet
in that range.  The observed superluminal motions place an upper limit
on the angle to the line-of-sight of about 20\arcdeg\ so the 120 pc
deprojects to at least 350 pc with much larger values more likely.

\item The outermost feature for which a speed could be determined is
the bulge at around 200 mas (Figure~\ref{medium18}).  It is moving at
a rate of 1.7 \masr, which is slower than most of the features closer
to the core, especially those with reliable speed measurements.  This
may be the leading edge of material ejected during a period of
enhanced activity.  It is now plowing into the older jet material plus
expanding the jet, a situation similar to that seen in 3C~84
\citep{W94}.  Such features in radio sources may constitute a fossil
record of previous periods of enhanced activity, about 100 years ago
in this case.

\item The 1.7 GHz data show evidence that there are both moving and
stationary features in the same regions (Figure~\ref{small18}).  In
fact, it appears that moving components can pass through a stationary
feature, dimming or brightening as they do so.  The stationary
features could either be places of interaction with the external
medium or could reflect changes in jet direction with corresponding
changes in the intensity of the relativistic beaming.

\item The stationary brightening regions have structure reminiscent of
the feature at 4\arcsec\ which also appears to be stationary to
within the measurement errors \citep{W97}.  This enhances the suspicion
that the 4\arcsec\ knot's lack of motion does not imply that the
underlying jet is not relativistic.  The knot may, like the stationary
brightening regions in the 1.7 GHz images here, be regions of interaction
with the external medium or a result of jet bending, perhaps as part of
a helical pattern.  The lack of a visible counterjet, despite the 
presence of a counter lobe on scales of several arcminutes, is further
evidence that the jet remains relativistic to large scales.

\item Some of the features seen in the 1.7 GHz images suggest that the
jet has a helical structure.  To match the observed structure, either
the jet must be extremely close to the line-of-sight (much closer than
the maximum angle allowed by the superluminal motion alone) or there
must be an offset between the line traced by the peak brightness and
the path of individual fluid elements.  The fluid equations (e.g.,
Hardee 2000) applied to helically twisted jets predict such an offset
because the region of high pressure describes a helix of larger
transverse amplitude than is followed by the fluid elements.
Additional theoretical constraints related to the low pattern speed
suggest that the jet is closer to the line-of-sight than about
$10\arcdeg$.  For any line-of-sight angle allowed by the superluminal
motion, the wavelength of the helix is relatively long, in contrast to
the situation in M87 where tightly wound helical patterns are seen.
Longer wavelengths are favored when the jet is surrounded by a
relatively cool medium.

\end{itemize}

   The picture that emerges is one of a continuous, relativistic jet
with much internal structure but a smooth outer envelope.  The
velocity is maintained from sub-parsec to hundreds of parsec scales.
There is some slowing at the largest scale feature whose speed could
be measured, but the jet appears to remain relativistic to kiloparsec
and larger scales.  The structure includes both features that move
with the jet, perhaps density enhancements or shocks, and also
features that do not move with the flow.  The moving features can pass
through the stationary features, suggesting that the stationary
features do not significantly disrupt the flow.  These could be places
where there is some interaction with the external medium or they could
be the result of variations in the jet path.  Individual features are
followed for large fractional distances from the core, suggesting that
they represent variations in fundamental and enduring parameters of
the jet such as density.  Possible helical patterns of long wavelength
are seen on several scales.  Theoretical arguments suggest
such wavelengths when the jet is cool and/or of high Lorentz factor
and is moving through a cool and not relativistically hot medium.

\acknowledgements 

We would like to thank the staffs of all the observatories listed in
the Table~\ref{antennas} and the Caltech, Charlottesville, and Socorro
correlators, without whose efforts these observations would not have
been possible.  While we did not use their data directly in the above
text, we often consulted the University of Michigan Radio Astronomy
Observatory flux density history available on the Internet for a
consistent history of activity in 3C~120.  That project is supported
by funds from the University of Michigan.  T. Hunter and M. Lystrup
acknowledge support as summer students at NRAO by the Research
Experiences for Undergraduates program of the National Science
Foundation.  P. Hardee acknowledges support from the National Science
Foundation through grant AST-9802955 to the University of Alabama.
P. Hardee and C. Walker acknowledge the Aspen Center for Physics where
some of this work was performed.

\clearpage



\clearpage

\begin{deluxetable}{lllr}
\tablenum{1}
\tablewidth{0pt}
\tablecaption{VLBI Antennas\label{antennas}}
\tablehead{
  \colhead{Name} & 
  \colhead{Code} &
  \colhead{Location} &
  \colhead{Diameter} \\
  \colhead{}& 
  \colhead{} &
  \colhead{} &
  \colhead{(m)}
  }  
\startdata
US VLBI Network: \\
~~~Arecibo              & A   & Puerto Rico     &300 \\
~~~Haystack             & K   & Massachusetts   & 37 \\
~~~Green Bank 140'      & G   & West Virginia   & 43 \\
~~~North Liberty        & I   & Iowa            & 18 \\
~~~Fort Davis (GRAS)    & F   & Texas           & 26 \\
~~~VLA (single antenna) & Y1  & New Mexico      & 25 \\
~~~VLA (phased array)   & Yp  & New Mexico      & $27 \times 25$ \\
~~~Owens Valley         & O   & California      & 40 \\
~~~Hat Creek            & H   & California      & 26 \\
~~~Maryland Point (NRL) & N   & Maryland        & 26 \\
European VLBI Network: \\
~~~Jodrell Mark I       & J   & United Kingdom  & 76 \\
~~~Jodrell Mark II      & J2  & United Kingdom  & 38 by 25 \\
~~~Onsala               & S   & Sweden          & 26 \\
~~~Westerbork (1 or 12 antennas)  & W   & The Netherlands & ($12 \times$) 25 \\
~~~Effelsberg           & B   & Germany         & 100 \\
~~~Medicina             & M   & Italy           & 32 \\
~~~Torun 18             & T   & Poland          & 18 \\
~~~Torun 32             & Tr  & Poland          & 32 \\
~~~Simeiz               & R   & Ukraine          & 22 \\
~~~Noto                 & Nt  & Italy           & 32 \\
VLBA: \\  
~~~Saint Croix          & Sc  & Virgin Islands  & 25 \\ 
~~~Hancock              & Hn  & New Hampshire   & 25 \\ 
~~~North Liberty        & Nl  & Iowa            & 25 \\ 
~~~Fort Davis           & Fd  & Texas           & 25 \\ 
~~~Los Alamos           & La  & New Mexico      & 25 \\ 
~~~Pie Town             & Pt  & New Mexico      & 25 \\
~~~Kitt Peak            & Kp  & Arizona         & 25 \\ 
~~~Owens Valley         & Ov  & California      & 25 \\ 
~~~Brewster             & Br  & Washington      & 25 \\ 
~~~Mauna Kea            & Mk  & Hawaii          & 25 \\ 
Other: \\
~~~Defford              & U   & United Kingdom  & 25 \\
~~~Cambridge            & E   & United Kingdom  & 18 \\
~~~Dwingeloo            & D   & The Netherlands & 25 \\
~~~Penticton            & P   & British Columbia & 25 \\
~~~Pushkino             & Pu  & Russia          & 25 \\
~~~Hartebeesthoek       & X   & South Africa    & 26 \\
\enddata
\end{deluxetable}

\begin{deluxetable}{crlr}
\tablenum{2}
\tablewidth{0pt}
\tablecaption{The 3C~120 VLBI Observations at 5.0 and 10.7 GHz \label{obs6}}
\tablehead{
  \colhead{Epoch} & 
  \colhead{Frequency} &
  \colhead{Antennas \tablenotemark{a}} &
  \colhead{Image Peak} \\
  \colhead{}& 
  \colhead{(GHz)} &
  \colhead{} &
  \colhead{(\jb)}
  }  

\startdata
   1978.91 &  10.7 &  KGFO           & 2.07 \\                    
   1979.25 &   5.0 &  BGFOH          & 1.30 \\			  
   1979.44 &  10.7 &  BKGFO          & 0.84 \\			  
   1979.92 &   5.0 &  BKGFO          & 0.85 \\			  
   1980.27 &  10.7 &  BKGO           & 1.18 \\			  
   1980.52 &  10.7 &  BKGFO          & 1.22 \\			  
   1980.72 &   5.0 &  BKFO           & 1.51 \\			  
   1981.10 &  10.7 &  KGFO           & 0.84 \\			  
   1981.63 &   5.0 &  BKGYpOH        & 0.88 \\			  
   1981.93 &   5.0 &  BSKGFYpOH      & 1.14 \\			  
   1982.25 &   5.0 &  BSAGFY1OH      & 1.21 \\			  
   1982.58 &   5.0 &  BKGFYpO        & 0.92 \\			  
   1982.94 &   5.0 &  BSWJAKNGFYpO   & 0.83 \\			  
   1983.25 &   5.0 &  BAKGO          & 0.45 \tablenotemark{b} \\ 
   1983.57 &   5.0 &  BSKNGFYpO      & 0.64 \\			  
   1983.91 &   5.0 &  BSKGIFYpO      & 0.62 \\			  
   1984.25 &   5.0 &  BSKGFY1OH      & 0.53 \\			  
   1984.40 &   5.0 &  SMAKGFYpOH     & 0.48 \\			  
   1984.78 &   5.0 &  BSKGFYpOK      & 0.78 \\			  
   1985.12 &   5.0 &  BSAKGIFY1O     & 1.65 \\			  
   1985.41 &   5.0 &  SAKNGIY1OH     & 2.31 \\			  
   1986.42 &   5.0 &  BSAKGFY1OH     & 0.89 \\			  
   1986.90 &   5.0 &  BSKGY1O        & 0.89 \\			  
   1987.41 &   5.0 &  BSWJMTXKGY1OH  & 0.84 \\			  
   1987.73 &   5.0 &  BSAKNGIFY1OH   & 0.68 \\			  
   1988.19 &   5.0 &  BSWJAKGIFY1O   & 0.56 \\			  
   1988.45 &   5.0 &  BSWJMKNPtY1O   & 0.50 \\			  
   1988.87 &   5.0 &  BSJMKGIFPtO    & 0.59 \\			  
\enddata
\tablenotetext{a}{See Table~\ref{antennas} for station codes.}
\tablenotetext{b}{Amplitude calibration suspect for this epoch --- see text.}
\end{deluxetable}

\begin{deluxetable}{cccc}
\tablenum{3}
\tablewidth{0pt}
\tablecaption{The 3C~120 VLBI Observations at 1.7 GHz \label{obs18}}
\tablehead{
  \colhead{Epoch} & 
  \colhead{Antennas \tablenotemark{a}} &
  \colhead{Image Peak \tablenotemark{b}} &
  \colhead{Image RMS \tablenotemark{b}} \\
  \colhead{} & 
  \colhead{} &
  \colhead{(\jb)} &
  \colhead{(m\jb)}
  }  
\startdata
   1982.78  & S B D J K N G A I F YpO P H            & 2.46 & 0.25  \\ 
   1984.26  & S B W J2T E U R K N G A I F YpO P H    & 0.61 & 0.17  \\  
   1989.85  & S B W R PuM X J A K G I PtKpLaY1       & 1.50 & 0.21  \\  
   1994.44  & VLBA\tablenotemark{c}                  & 0.58 & 0.12  \\
   1997.71  & B J M NtTrW G Y1 VLBA\tablenotemark{c} & 1.05 & 0.05  \\
\enddata
\tablenotetext{a}{See Table 1 for station codes.}
\tablenotetext{b}{For the images in Figures~\ref{medium18} and 
\ref{small18}.}
\tablenotetext{c}{``VLBA'' means all 10 VLBA antennas.}
\end{deluxetable}

\begin{deluxetable}{ccrrr}
\tablenum{4}
\tablewidth{0pt}
\tablecaption{Fitted Component Speeds for Features Seen at 5 GHz
\label{speeds}}
\tablehead{
  \colhead{Feature} & 
  \colhead{Number of} &
  \colhead{Speed\tablenotemark{a}} &
  \colhead{Start} &
  \colhead{$v/c$} \\
  \colhead{} & 
  \colhead{Epochs} &
  \colhead{(\masr)} &
  \colhead{(date)} &
     }  
\startdata
 A & \phn4 &  $  1.47 \pm  0.46 $ &  $  1978.22 \;_{ -0.52}^{ +0.28}$ &   3.00 \\
 B &    13 &  $  3.14 \pm  0.12 $ &  $  1980.38 \;_{ -0.07}^{ +0.07}$ &   6.39 \\
 C & \phn5 &  $  2.17 \pm  0.43 $ &  $  1980.73 \;_{ -0.40}^{ +0.27}$ &   4.41 \\
 D &    10 &  $  2.72 \pm  0.16 $ &  $  1981.86 \;_{ -0.10}^{ +0.08}$ &   5.53 \\
 E & \phn6 &  $  2.40 \pm  0.47 $ &  $  1982.56 \;_{ -0.27}^{ +0.19}$ &   4.88 \\
 F & \phn5 &  $  2.41 \pm  0.75 $ &  $  1983.31 \;_{ -0.64}^{ +0.34}$ &   4.90 \\
 G & \phn3 &  $  2.42 \pm  1.70 $ &  $  1984.27 \;_{ -2.26}^{ +0.43}$ &   4.93 \\
 H & \phn3 &  $  3.37 \pm  0.58 $ &  $  1985.74 \;_{ -0.25}^{ +0.18}$ &   6.86 \\
 I & \phn7 &  $  3.20 \pm  0.18 $ &  $  1986.10 \;_{ -0.11}^{ +0.10}$ &   6.51 \\
 J & \phn6 &  $  2.23 \pm  0.27 $ &  $  1986.59 \;_{ -0.21}^{ +0.17}$ &   4.53 \\
 K & \phn3 &  $  1.45 \pm  1.09 $ &  $  1987.56 \;_{ -3.23}^{ +0.49}$ &   2.96 \\
\enddata
\tablenotetext{a}{The $v/c$ values assume $H_0 = 75$ \kms Mpc$^{-1}$.}
\end{deluxetable}

\begin{deluxetable}{p{0.32\textwidth}p{0.62\textwidth}}
\tablenum{5}
\tablewidth{\textwidth}
\tablecaption{Summary of Models Discussed and Their Implications
\parskip=40pt
\label{models}}
\tablehead{
  \colhead{Model} & 
  \colhead{Implications Given the Observational Constraints}
  }  

\startdata
Simple morphological model with the particles following the apparent helix.
&
Jet axis is very close to the line-of-sight or the brightness
variations would be too large.

\smallskip
Gamma is significantly higher than the mimimum required for 
the observed superluminal motion.
\\[15pt]

Physical theory focusing on the helix wavelength and the pressure 
distribution.
&
Long helical wavelength implies cool external medium.

\smallskip
High pressure regions and, hence, the brightest emission 
is to the outside of jet.

\smallskip
Apparent helix from the brightest emission can have a
larger radius than the helical paths followed by the particles.
\\[15pt]

Simple morphological model with the apparent helix wider than the particle 
path helix. 
&
Allows wider angles to line of sight without excessive brightness
variations.

\smallskip
Gamma is near the normal value for the observed superluminal motion.
\\[15pt]
Physical theory focusing on the pattern speed. &
  Low observed pattern speed also implies cool external medium.

\smallskip
  Moving pattern stretches apparent helix wavelength.

\smallskip
  Transverse velocities are not as high as wavelength suggests.

\smallskip
  Constrains angle to line-of-sight to be under about $10\arcdeg$.
\\

\enddata
\end{deluxetable}

\end{document}